\documentclass[structabstract]{aa} 
\usepackage{longtable}
\usepackage[tableposition=below]{caption}
\usepackage{tablefootnote}
\usepackage{lscape}
\usepackage{graphicx}
\usepackage{epstopdf}
\usepackage{txfonts}

\begin{document} 

\title{Three-dimensional simulations of the interaction between the nova ejecta, the accretion disk, and the companion star}

\author{Joana Figueira \inst{1,2}
   \and Jordi Jos\'e \inst{1,2}
   \and Enrique Garc\'\i a-Berro \inst{2,3}
   \and Simon W. Campbell \inst{4,5,6} 
   \and Domingo Garc\'\i a-Senz \inst{1,2}
   \and Shazrene Mohamed \inst{7,8,9} }

\offprints{J. Jos\'e}

 \institute{Departament de F\'\i sica, EEBE,
            Universitat Polit\`ecnica de Catalunya,
            c/Eduard Maristany 10,
            E-08930 Barcelona,
            Spain\
        \and
            Institut d'Estudis Espacials de Catalunya,
            c/Gran Capit\`a 2-4,
            Ed. Nexus-201,
            E-08034 Barcelona,
            Spain\
        \and
            Departament de F\'\i sica,
            Universitat Polit\`ecnica de Catalunya,
            c/Esteve Terrades 5,
            E-08860 Castelldefels,
            Spain\
        \and
            Max-Planck-Institut f\"ur Astrophysik, Karl-Schwarzschild-Strasse 1, 
            D-85748 Garching bei M\"unchen, Germany
        \and
            School of Physics and Astronomy, Monash University, Clayton
            3800, Victoria, Australia
        \and
             Monash Centre for Astrophysics (MoCA), Monash University,
             Clayton 3800, Victoria, Australia
        \and
            South African Astronomical Observatory, PO Box 9, Observatory Rd., 7935 Cape Town, 
            South Africa 
        \and
            Astronomy Department, University of Cape Town, 7701 Rodenbosch, 
            South Africa 
        \and
            South Africa National Institute for Theoretical Physics, Private Bag X1, Matieland,
            7602 Stellenbosch, South Africa 
            \\
      \email{jordi.jose@upc.edu}}

\date{\today}

\abstract{Classical novae  are thermonuclear explosions hosted by accreting white dwarfs in stellar binary  systems. Material   piles up  on top of the white dwarf star under  mildly  degenerate   conditions, driving a thermonuclear  runaway. The energy released by the suite of  nuclear processes operating at  the envelope (mostly proton-capture reactions and $\beta^+$-decays) heats the material up to peak temperatures ranging from 100 to 400 MK. In these events,  about $10^{-3} -  10 ^{-7}$ $M_{\odot}$,  enriched  in  CNO  and, sometimes,  other  intermediate-mass  elements (e.g., Ne, Na, Mg, Al), are  ejected  into the interstellar  medium.}
{To date, most of the efforts undertaken in the modeling of
classical nova outbursts have focused on the early stages of the explosion
and ejection, ignoring the interaction of the ejecta, first with the accretion disk orbiting the
white dwarf, and ultimately with the secondary star.}
{A suite of three-dimensional, SPH simulations of the interaction between the nova ejecta, the accretion disk, and the stellar companion have been performed to fill this gap,  aimed at testing the influence of the different parameters (i.e., mass and velocity
of the ejecta, mass and geometry of the accretion disk) on the dynamical and chemical properties of the system.}
{We discuss the conditions that lead to the disruption of the accretion disk and to mass loss from the binary system. In addition, we discuss the likelihood of chemical contamination of the stellar secondary induced by the impact with the nova ejecta and its potential effect on the next nova cycle.}
         {}

\keywords{(Stars:) novae, cataclysmic variables --- nuclear reactions, nucleosynthesis, abundances --- hydrodynamics}

\titlerunning{3D simulations of the interaction between the nova ejecta and the binary system}
\authorrunning{J. Figueira et al.} 

\maketitle

\section{Introduction}
The coupling of spectroscopic determinations of chemical abundances, 
photometric studies of light curves, and hydrodynamic simulations of the
accretion, expansion and ejection stages, 
has been instrumental in our understanding of the nova phenomenon. 
The scenario envisaged assumes a white dwarf star hosting the explosion in a
close binary system (see, e.g., Sanford 1949, Joy 1954, and Kraft 1964, for 
some of the first systematic observations that revealed the binary nature of novae).
 The low-mass, main sequence stellar companion (frequently, a K-M dwarf, 
although observations increasingly support the presence of more evolved 
companions in some systems) overfills its Roche lobe, and matter flows 
through the inner Lagrangian point of the system. 
 A fraction of this hydrogen-rich matter lost by the secondary spirals in via an accretion disk and 
ultimately piles up on top of the white dwarf (typically, at a rate 
$\sim 10^{-8} - 10^{-10}$ $M_\odot$ yr$^{-1}$). The accreted envelope layers get 
gradually compressed by the continuous matter infall  under mildly
degenerate conditions. Compressional heating  initiates nuclear reactions 
and a thermonuclear runaway ensues. The thermonuclear 
origin of nova explosions was first theorized by Schatzman (1949, 1951). 
This was followed by a number of significant contributions in the 1950s and 
1960s (see, e.g., Cameron 1959, Gurevitch \& Lebedinsky 1957), 
including pioneering attempts to mimic the explosion through the coupling 
of radiative transfer in an optically thick expanding shell with hydrodynamics
  (Giannone \& Weigert 1967, Rose 1968, Sparks 1969). 

To date, most of the efforts undertaken in the modeling of nova outbursts 
have focused on the early stages of the explosion and ejection (see Starrfield, Iliadis 
\& Hix 2008, 2016, Jos\'e \& Shore 2008, and Jos\'e 2016, for recent reviews).
Therefore, key aspects of the evolution of these systems, associated with the 
interaction of the ejecta, first with the accretion disk orbiting the 
white dwarf, and ultimately with the secondary star, have been largely 
unexplored\footnote{Note, however, that a similar scenario, the interaction between
the material ejected in a type Ia supernova  and the companion star, has been addressed in a number of papers
(see, e.g., Marietta, Burrows \& Fryxell 2000, and Garc\'\i a-Senz, Badenes \& Serichol 2012).}.
Shortly after the outer layers of the white dwarf expand and achieve escape 
velocity, a fraction of the ejected material is expected to collide 
with the secondary star. 
As a result, part of the nova ejecta will mix with the outermost layers of the 
secondary. 
The resulting chemical contamination may have 
potential implications for the next nova cycle, once mass transfer from the
secondary resumes.

   \begin{table*}[thb]
   \caption{Models computed.}
   \label{Series1}
   \centering                         
   \begin{tabular}{c c c c c c }   
   \hline\hline\
   Model& $M_{\rm ejecta}$& $M_{\rm disk}$& $V_{\rm ejecta}^{\rm max}$& $H/R$& 
                             Disk\\ 
        &      ($M_\odot$)&    ($M_\odot$)&              (km s$^{-1}$)&      & 
                                                              disruption\\ 
   \hline
   A & $5.14 \times 10^{-4}$ & $ 2.04 \times 10^{-5}$ & 1200 & $0.03$ &
       Yes\\ 
   $\rm A_{hres}$$^a$& $5.14 \times 10^{-4}$ & $ 2.04 \times 10^{-5}$ & 1200 & $0.03$ &
        Yes\\ 
   B & $5.14 \times 10^{-4}$ & $ 2.04 \times 10^{-5}$ & 800  & $0.03$ &
       Yes\\
   C & $5.14 \times 10^{-4}$ & $ 9.28 \times 10^{-5}$ & 1200 & $0.03$ &
       No\\ 
   D & $5.14 \times 10^{-4}$ & $ 9.28 \times 10^{-5}$ & 800  & $0.03$ &
       No\\ 
   E & $1.60 \times 10^{-3}$ & $2.04 \times 10^{-5}$ & 1200  & $0.03$ & 
       Yes\\ 
   F & $1.60 \times 10^{-3}$ & $2.04 \times 10^{-5}$ & 800   & $0.03$ & 
       Yes\\
   G & $5.14 \times 10^{-4}$ & $3.96 \times 10^{-5}$ & 1200  & $0.06$ & 
       No\\ 
   H & $1.60 \times 10^{-3}$ & $3.96 \times 10^{-5}$ & 1200  & $0.06$ & 
       Yes\\ 
   I & $5.14 \times 10^{-4}$ & $ 9.28 \times 10^{-5}$ & 3000 & $0.03$ &
       No\\ 
   J & $5.14 \times 10^{-4}$ & $3.96 \times 10^{-5}$ & 3000  & $0.06$ & 
       Yes\\ 
   K & $5.14 \times 10^{-4}$ & $ 2.04 \times 10^{-5}$ & 1200 & $0.06$ &
       Yes\\ 
    \hline
    \end{tabular}
\vspace{0.1 cm}
\begin{list}{}{}
\item[$^{\mathrm{a}}$] Model computed with twice the number of particles than Model A.
\end{list}
    \end{table*}

Novae are also prolific dust producers. Infrared
(Evans \& Rawlings 2008, Gehrz 1998, 2008) and ultraviolet observations 
(Shore et al. 1994) have unambiguously revealed dust forming episodes
in the ejected shells accompanying some nova outbursts, about $\sim$ 100 days 
after the explosion. In fact, it has been suggested that novae may have 
contributed to the inventory of presolar grains isolated from 
meteorites. A major breakthrough in the identification of nova 
candidate grains was
achieved by Amari et al. (2001, 2002), who reported several SiC and 
graphite grains, isolated from the Murchison and Acfer 094 meteorites, 
with abundance patterns qualitatively similar to those predicted by 
models of  nova outbursts: low $^{12}$C/$^{13}$C and $^{14}$N/$^{15}$N 
ratios, high $^{30}$Si/$^{28}$Si and close-to-solar $^{29}$Si/$^{28}$Si
ratios, and high $^{26}$Al/$^{27}$Al and $^{22}$Ne/$^{20}$Ne ratios.
However, to quantitatively match the grain data, 
mixing between material synthesized
in the explosion and more than ten times as much unprocessed, 
isotopically close-to-solar material was required. The collision of the 
ejecta, either with the accretion disk or with the secondary star, 
may naturally provide the required chemical dilution to explain the reported
grain data. 

Moreover, the unexpected discovery of very high-energy emission ($>$ 100 MeV),
first observed in the symbiotic binary V407 Cygni (Abdo et al. 2010), and 
subsequently detected in a number of novae (e.g., V407 Cyg, V1324 Sco, V959 Mon, 
V339 Del, V1369 Cen), by the Large Area Telescope on board the Fermi 
$\gamma$-ray space observatory (Fermi LAT), has also been 
linked to shock acceleration in the ejected shells after interaction 
with a wind from the secondary\footnote{Diffusive shock acceleration of 
electrons and protons, with a maximum energy of a few TeV, was  
predicted by Tatischeff \& Hernanz (2007),  
 in the framework of the 2006 outburst of the recurrent nova RS Ophiuchi. 
See also Shore et al. (2013) for an explanation of the origin of 
X-ray emission in Nova Mon 2012 based on internal shocks driven by 
the collision of filaments that freeze out during expansion.}. 
This has confirmed novae as a distinct class of $\gamma$-ray 
sources (Ackermann et al. 2014).  

All the abovementioned aspects stress the need for a thorough 
description of the dynamics of the system after the nova explosion, 
following the collision of the ejecta with the accretion disk, and 
subsequently, with the secondary star. The present paper aims at filling  
this gap. 
The manuscript is organized as follows. The method of computation, the input 
physics, and the initial conditions adopted are described in Sect. 2. 
A full account of the different 3D simulations of the interaction of the 
ejecta with the accretion disk, and ultimately with the main sequence 
companion, is presented in Sect. 3. 
 The effect of the different parameters on the stability of the accretion
disk as well as on the amount of mass lost from the system is also analyzed
in Sect. 3.
 Discussion on the expected level of 
chemical contamination of the outer layers of the secondary star is given in Sect. 4.
A summary of the most relevant conclusions of this paper is presented in Sect. 5.

\section{Model and input physics}  
\subsection{Initial configuration}
The 3D computational domain\footnote{The presence of a disk, and its key role in the simulations reported 
in this paper, does not allow us to rely on SPH axisymmetric codes to increase resolution of the models, in constrast
to other astrophysical scenarios such as type Ia supernovae (see, e.g., Garc\'\i a-Senz, Badenes \& Serichol 2012).}
 of the simulations discussed in this paper includes the white dwarf that hosts the nova explosion, the expanding nova ejecta, the accretion disk, and the main sequence companion.

\subsubsection{The white dwarf star}
The white dwarf is modelled as a 0.6 $M_{\odot}$ point-like mass, 
which is enough to account for its gravitational pull on the system.
The 
expanding ejecta, which at the beginning of the 3D simulations is located 
between 0.65 $R_{\odot}$ (inner edge) and 0.72 $R_{\odot}$ (outer edge) 
from the underlying white dwarf, has a mean metallicity of $Z = 0.54$, 
and a mass, density and 
velocity profiles corresponding to the values computed with the 1D code {\tt SHIVA} (see Sect. 2.2, for details). 

\subsubsection{The main sequence star}
\begin{figure*}[bth]
\centering
\includegraphics[width=10 cm]{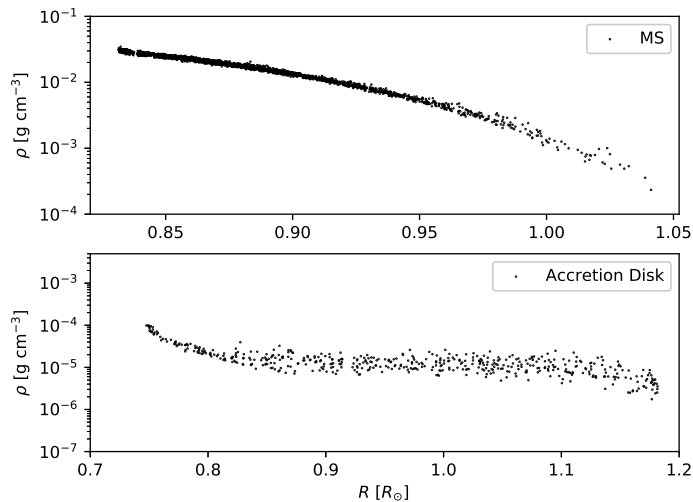}
\caption{Initial density profiles for the outer main 
sequence layers (after relaxation) and the accretion disk.} 
\label{FigDensity}
\end{figure*}

A 1 $M_\odot$, solar metallicity, main sequence companion is adopted as the 
secondary.  The star has spherical symmetry and is built
in hydrostatic equilibrium conditions. A polytropic equation of state with $\gamma=5/3$ has been
considered. The corresponding density and internal energy profiles were subsequently
mapped onto a 3D particle distribution. To avoid a substantial computational  load, only the 
outer layers of the main sequence star have been considered, since during the collision between
the nova ejecta and the secondary, particles are not expected to penetrate deep inside the 
star. Even though preliminary simulations performed in this work suggest that
the ejecta may penetrate, at most, through the outer $\sim$ 0.1 $R_{\odot}$ of the 
secondary, the outermost $\sim$ 0.2 $R_\odot$ (0.15 $M_\odot$) of the star have been taken into
account.  The  
rest of the star has been replaced by a point-like mass located at its center. 
To generate the initial 3D particle distribution of the outer main 
sequence layers, the initial 1D density profile is sliced into several shells of equal radius. 
For each shell, a `glass' technique has been implemented (White 1996): in essence, 
a cube is filled with a random number of particles until a uniform distribution is achieved, from which a shell with constant density 
is extracted. The same procedure is used for each shell at different densities and has also been adopted 
for the accretion disk and the ejecta (assuming in this case axial symmetry). 
The initial density profile for the outer main sequence layers is shown in Fig. ~\ref{FigDensity}.
For convenience, and to guarantee good accuracy in the interpolated functions, all SPH
particles used in this work have the same mass, $\sim 10^{-8}$
$M_\odot$.
One can easily infer the number of SPH particles within each mass
shell from its total mass. 
About 3.8 million SPH particles have been used to model the outer $\sim 0.2$
$R_\odot$ (0.15 $M_\odot$) of the secondary. 
To guarantee  that the resulting 3D structure is in  hydrostatic equilibrium, the stellar secondary is relaxed for a total time of the order
of 20 orbital periods.

\begin{table*}[t]
\caption{Mass (ejecta plus disk) gravitationally bound to the secondary star, 
    $\Delta M_{\rm MS}$, and to the white dwarf, 
    $\Delta M_{\rm WD}$, and total mass leaving the binary system, $\Delta M_{\rm esc}$, together with their 
    fractions (in \%) over the 
     total ejecta plus disk masses, after collision with the nova ejecta. $\Delta M_{\rm MS, lost}$ is the mass lost by the main sequence star in the interaction with the nova ejecta.
} 
    \label{Series1}
    \centering   
    \begin{tabular}{c c c c c c c c}
    \hline\hline\
    Model& $\Delta M_{\rm MS}$($M_{\odot}$) & $\Delta M_{\rm WD}$($M_{\odot}$) & $\Delta M_{\rm esc}$($M_\odot$)   &
    $\frac{\Delta M_{\rm MS}}{(M_{\rm ejecta} + M_{\rm disk})}$  & 
    $\frac{\Delta M_{\rm WD}}{(M_{\rm ejecta} + M_{\rm disk})}$  & 
    $\frac{\Delta M_{\rm esc}}{(M_{\rm ejecta} + M_{\rm disk})}$  
&    $\Delta M_{\rm MS, lost}$($M_{\odot}$) 
\\
    \hline
    A & $ 1.95 \times 10^{-5}$ & $4.64 \times 10^{-5}$ & $ 4.69\times 10^{-4}$ & 3.65\% & 8.68\% & 87.7\% &$2 \times 10^{-7}$\\
    $\rm A_{hres}$ & $ 2.05 \times 10^{-5}$ & $4.83 \times 10^{-5}$ & $ 4.66\times 10^{-4}$ & 3.83\% & 9.03\% & 87.2\% &$1.4 \times 10^{-7}$\\
    B & $ 2.92 \times 10^{-5}$ & $1.01 \times 10^{-4}$ & $ 4.05\times 10^{-4}$ & 5.46\% & 18.9\% & 75.6\% & -\\
    C & $ 4.96 \times 10^{-5}$ & $1.50 \times 10^{-4}$ & $ 4.08\times 10^{-4}$ & 8.17\% & 24.7\% & 67.2\% &$1.6 \times 10^{-7}$\\
    D & $ 6.20 \times 10^{-5}$ & $1.83 \times 10^{-4}$ & $ 3.62\times 10^{-4}$ & 10.2\% & 30.2\% & 59.7\% &$1.2 \times 10^{-7}$\\
    E & $ 5.62 \times 10^{-5}$ & $1.12 \times 10^{-4}$ & $ 1.45\times 10^{-3}$ & 3.47\% & 6.93\% & 89.7\% &$1.4 \times 10^{-6}$\\
    F & $ 9.82 \times 10^{-5}$ & $2.66 \times 10^{-4}$ & $ 1.26\times 10^{-3}$ & 6.06\% & 16.4\% & 77.6\% &$4.8 \times 10^{-7}$\\
    G & $ 5.69 \times 10^{-5}$ & $1.71 \times 10^{-4}$ & $ 3.26\times 10^{-4}$ & 10.3\% & 30.9\% & 58.9\% &$4 \times 10^{-8}$\\
    H & $ 6.06 \times 10^{-5}$ & $1.25 \times 10^{-4}$ & $ 1.46\times 10^{-3}$ & 3.69\% & 7.60\% & 88.8\% &$1.3 \times 10^{-6}$\\
    I & $ 2.17 \times 10^{-5}$ & $4.79 \times 10^{-5}$ & $ 5.40\times 10^{-4}$ & 3.57\% & 7.88\% & 88.9\% &$2.4 \times 10^{-6}$\\
    J & $ 1.66 \times 10^{-5}$ & $5.55 \times 10^{-5}$ & $ 4.83\times 10^{-4}$ & 3.00\% & 10.0\% & 87.2\% &$1.4 \times 10^{-6}$\\
    K & $ 5.39 \times 10^{-5}$ & $1.61 \times 10^{-4}$ & $ 3.20\times 10^{-4}$ & 10.1\% & 30.1\% & 59.8\% & -\\
    \hline
    \end{tabular}
    \end{table*}

\subsubsection{The mass-accretion disk}
The accretion disk that orbits the point-like white dwarf in Keplerian 
rotation is modeled according to the Shakura-Sunyaev, V-shaped disk solution 
(Shakura \& Sunyaev 1973, Frank, King \& Raine 2002). 
In the fiducial model reported in this paper (hereafter, Model A; see Table 1)
a solar-composition disk, with a mass of $2 \times 10^{-5}$ $M_\odot$, and a geometry 
given by a ratio of height to radius of $H/R = 0.03$, 
has been assumed  
(see Sects. 3.4 and 3.5, for the effect of these parameters on the simulations). 
Other models of accretion disks (e.g., flared disks) and inclusion of alternative assumptions (smaller 
extended disks; see Warner 2003) will be addressed in a future paper ---see 
Puebla, Diaz \& Hubeny (2007), for a comparison between current disk models and observational data.
In this work, the accretion disk contains only a few thousand SPH particles, 
being the truly limiting factor of the simulations (the nova ejecta contains up to 19,000 SPH particles).
The sound-crossing time throughout the disk is $\sim$ 4 hours, while the time 
required for the ejecta to reach and hit the disk is $\sim$ 6 minutes. Therefore, no relaxation
of the disk, also built by means of the cubic `glass' technique, is needed.
The initial density profile for the mass-accretion disk is shown in Fig. ~\ref{FigDensity}.

\subsection{Model}

The first stages of the evolution of the nova outbursts, through accretion, expansion,
and ejection, have been modeled by means of the 1D, spherically 
symmetric, Lagrangian, hydrodynamic code {\tt SHIVA} (see Jos\'e \& 
Hernanz 1998, and Jos\'e 2016, for details).
When the inner edge of the ejecta reached a size of 0.65 $R_{\odot}$, the structure 
was mapped onto a 3D domain, that included as well the white dwarf that hosts the nova explosion, the accretion disk, and the main sequence companion. 
The evolution of the
 system was subsequently followed with the 3D smoothed-particle 
hydrodynamics (SPH) code {\tt GADGET-2} (Springel, Yoshida \& White 2001,
 Springel \& Hernquist 2002, 
Springel 2005). This parallelized, explicit, Lagrangian, 
mesh-free code describes fluids in terms of a set of discrete elements 
(hereafter, particles). Their continuous properties (e.g., density, temperature, velocity) are obtained 
through kernel interpolation and summation over all neighboring particles.
In the simulations presented in this paper, a cubic spline kernel has been adopted. 
To handle shocks, {\tt GADGET-2} uses the artificial viscosity 
prescription developed by Monaghan (1997), together with a viscosity-limiter 
for pure shear flows (Balsara 1995). 
In addition, the code computes gravitational interactions using a hierarchical
 oct-tree algorithm (Barnes \& Hut 1986).  
Time steps are controlled by means of a Courant factor taken as 0.15. 
The gravitational softening of the SPH particles has been approximated by their 
 smoothing lengths. 
It is worth noting that {\tt GADGET-2} enables density contrasts since it 
offers adaptive smoothing lengths (Springel \& Hernquist 2002). 

The characteristic size of the overall binary system, for a given set of values of the
masses of the primary and secondary stars, is determined by 
the orbital period. In the simulations reported in this work, a value of 
 $P_{\rm orb} = 8.9$ hr has been adopted. This is a representative value
of long-period classical nova systems,  $P_{\rm orb} >$ 7 hr, that may represent
about 20\% of all novae (Tappert et al. 2013).
Rotation of the binary system has not been considered, since the overall duration of
the interaction between the ejecta, the disk and the secondary is very small.   
In fact, the orbital period adopted is  $\sim 12$ times longer than 
the time it takes for the ejecta to reach and hit the main sequence companion.
 The effect of the Coriolis and centrifugal forces 
has also not been considered.
They introduce  angular momentum and viscous-shear dissipation,
which may affect the trajectories of some of the ejecta particles 
as they travel toward the secondary. 
Such effects will be addressed in a future
manuscript.

\begin{figure*}[bth]
\centering
\includegraphics[width=\textwidth]{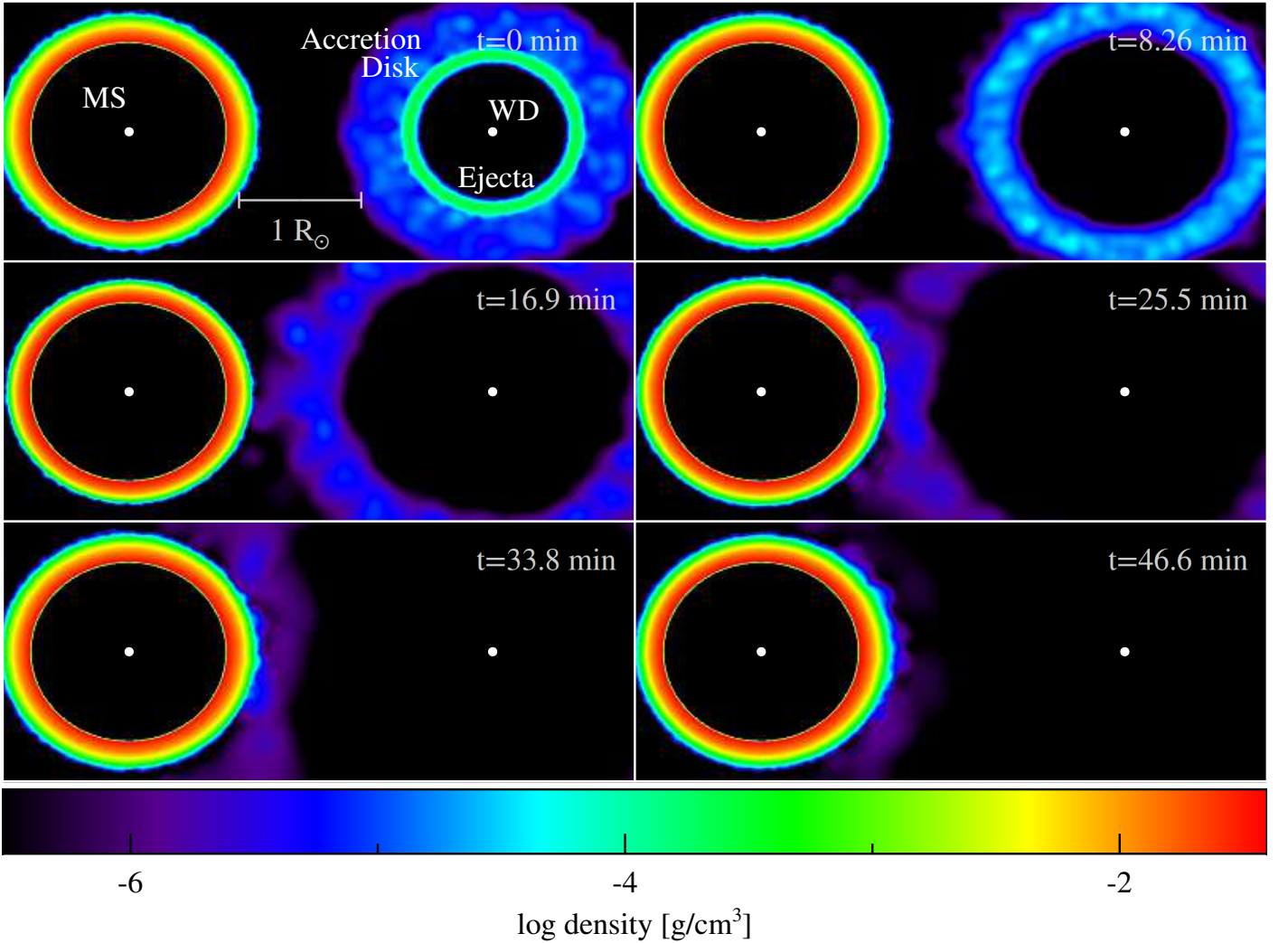}
\caption{Cross-sectional slice in the binary orbital plane (XY) showing the density of Model A
at different stages of the interaction between the nova ejecta 
and the accretion disk, subsequently followed by a collision with the main 
sequence companion. 
A movie showing the full evolution of
this model, {\tt modelA-XY.mov}, is available online and at {\tt http://www.fen.upc.edu/users/jjose/Downloads.html}.
See also {\tt modelA-YZ.mov}, for a movie depicting the evolution of the system from a side view (YZ plane). 
Snapshots and movies have been generated by means of the visualisation 
software SPLASH (Price 2007).}
\label{evolutionA}
\end{figure*}

\begin{figure*}[bth]
\centering
\includegraphics[width=10 cm]{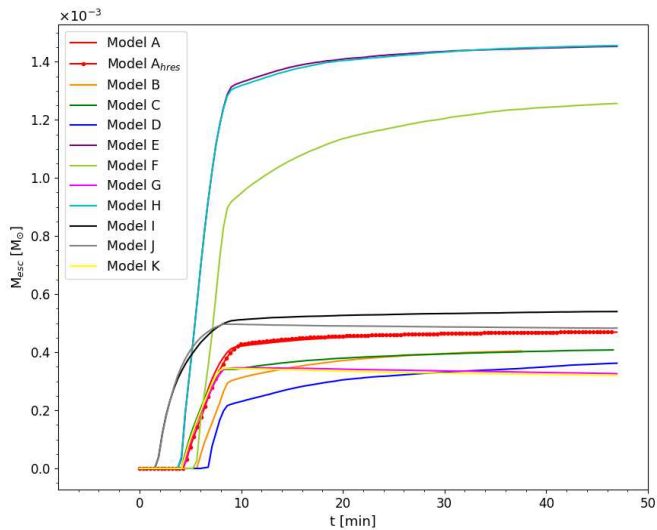}
\caption{Time evolution of the mass leaving the binary system for the different models reported in this paper.} 
\label{FigMloss}
\end{figure*}

\section{Results: Disk stability and mass loss} 
Observations suggest that the accretion disk does not always get disrupted by a nova outburst\footnote{See, however, Drake \& Orlando (2010), for simulations of recurrent nova systems leading always to full disruption of the accretion disks.}. Indeed, in some systems, the presence of a disk has been confirmed only a few months/years after the explosion (see, e.g., Leibowitz, Mendelson, \& Mashal 1992, Retter, Leibowitz, \& Kovo-Kariti 1998,      Retter, Leibowitz, \& Ofek 1997, Skillman et al. 1997, 
Hernanz \& Sala 2002), which is clearly at odds with the typical timescales required for a disk to assemble (of the order of decades).
It has been suggested that the accretion disk is only disrupted if the system is an intermediate polar (Retter 2003). In those systems, the magnetic field could truncate the inner regions
of the disk, which would be less massive than in non-magnetic systems, and therefore, prone to be disrupted by a nova explosion. 
However, it is worth noting that the mass and the mean density of such disks are also poorly constrained quantities from an observational viewpoint. 

To elucidate the possible effect of the nova outburst on the accretion disk, a suite of models aimed at testing the influence of the different parameters of the system (i.e., 
mass and velocity of the ejecta, mass and geometry of the accretion disk) have been considered (see Table 1). 

\subsection{Evolution of Model A}
Model A describes the interaction between $M_{\rm ejecta} =
5.1 \times 10^{-4}$ $M_\odot$, ejected from a 0.6 $M_\odot$ white dwarf during a nova
outburst (with a maximum velocity of the ejecta of $V_{\rm ejecta}^{\rm max} = 1200$ km s$^{-1}$), and a $2 \times 10^{-5}$ $M_\odot$ accretion disk (with $H/R$ = 0.03), subsequently 
followed by the collision with a 1 $M_\odot$ main sequence companion (see Table 1). 
Snapshots of the evolution of this model, in terms of density,  are displayed in Fig. ~\ref{evolutionA}. 
Movies showing the full evolution of
this model are available online and at {\tt http://www.fen.upc.edu/users/jjose/Downloads.html}.

The ejecta hits the disk a few seconds after the beginning of the simulation (Fig. ~\ref{evolutionA}, upper panels). 
The energy released during the collision heats the disk, which achieves a 
maximum temperature of $\langle T_{\rm disk}^{\rm max} \rangle \sim 1.4 \times 10^6$ K, with only a handful of SPH particles ($\sim 20$) reaching $T^{\rm max} \sim 1.2 
\times 10^7$ K. This suggests that nuclear reactions do not play a relevant role in the interaction\footnote{The same conclusion applies 
to all models reported in this paper.}. 
Only a small fraction of the nova ejecta hits the disk ($m'_{\rm ejecta} \sim 1\% M_{\rm ejecta}$), with a mean kinetic energy, $K = \frac{1}{2} m'_{\rm ejecta} V_{\rm ejecta}^2 
\sim 7 \times 10^{43}$ ergs. A crude estimate of the gravitational binding energy of the disk can be obtained from  $U \sim G M_{\rm WD} M_{\rm disk} / r_{\rm mean}$, where 
$G$ is the gravitational constant, $M_{\rm WD}$ is the mass of the underlying white dwarf, and $M_{\rm disk}$ and $r_{\rm mean}$ are the mass of the disk and the mean distance between
the white dwarf and the disk. Estimates for Model A yield $U \sim 6 \times 10^{43}$ ergs, which is similar to the kinetic energy of the impinging ejecta.
In Model A, simulations reveal the total disruption of the disk (middle left panel), which gets totally swept up and  mixed with the ejecta. 
However, other models with different choices for the geometry ($H/R$), mass, and velocity of the disk, and mass of the ejecta, may yield different outcomes (see below).

At about $t \sim 17$ min (middle panels), a mixture of ejecta and disk material impinges on the main
sequence companion. The temperature increases slightly in the outermost layers of the secondary, but not enough to spark nuclear reactions. 
In the collision, a subset of the ejecta/disk particles penetrate through the outer layers of the secondary, reaching a maximum depth of $\sim 1.1 \times 10^{-5}$ $M_{\odot}$ from the surface. 
The energy released in the collision drives a moderate expansion of the outer layers of the 
star (lower panels). Since the secondary overfills its Roche lobe, part of the material incorporated into the main sequence star will be later 
re-accreted by the white dwarf, as soon as mass-transfer resumes and the accretion disk is re-established.

About $\sim 4.7 \times 10^{-4}$ $M_\odot$ (i.e., 88\% of the mixture of disk and nova ejecta) leave the binary system in Model A. In contrast, only $\sim 2 \times 10^{-5}$ $M_\odot$ 
(3.7\%; mostly nova ejecta) remain gravitationally bound to the main sequence companion, while $\sim 4.6 \times 10^{-5}$ $M_\odot$ (8.7\%) are bound to the white dwarf (see Table 2). 
A small amount of mass, $2 \times 10^{-7}$ $M_\odot$, involving only a handful of SPH particles, is expelled from the outer main sequence layers in the interaction with the nova ejecta. Figure ~\ref{FigMloss} shows the time evolution of the mass leaving the binary system in Model A, and for all models reported in this paper. 
The early and sharp increase in mass loss ($t \leq 10$ min) results from the interaction between the nova ejecta and the disk, when the latter gets totally swept up and mixed with the former. Most of the ejecta and disk mixture leaves the binary system. The longer-term evolution of the mass loss plot 
($t > 10$ min) reveals that little is expelled from the outer main sequence layers as a result of the impact with the nova ejecta.

 To test the feasibility of these results, a higher resolution run with twice the number of 
particles than Model A (hereafter, Model $\rm A_{hres}$) was also performed. As shown in Table 2 (see also Fig. 3), 
both models A and $\rm A_{hres}$
yield similar results, which suggests that the overall number of particles adopted in this paper was
appropriate.  Movies showing the full evolution of
Model $\rm A_{hres}$ are also available online and at {\tt http://www.fen.upc.edu/users/jjose/Downloads.html}.

\subsection{Effect of the mass of the nova ejecta}
\begin{figure*}[thb]
\centering
\includegraphics[width=\textwidth]{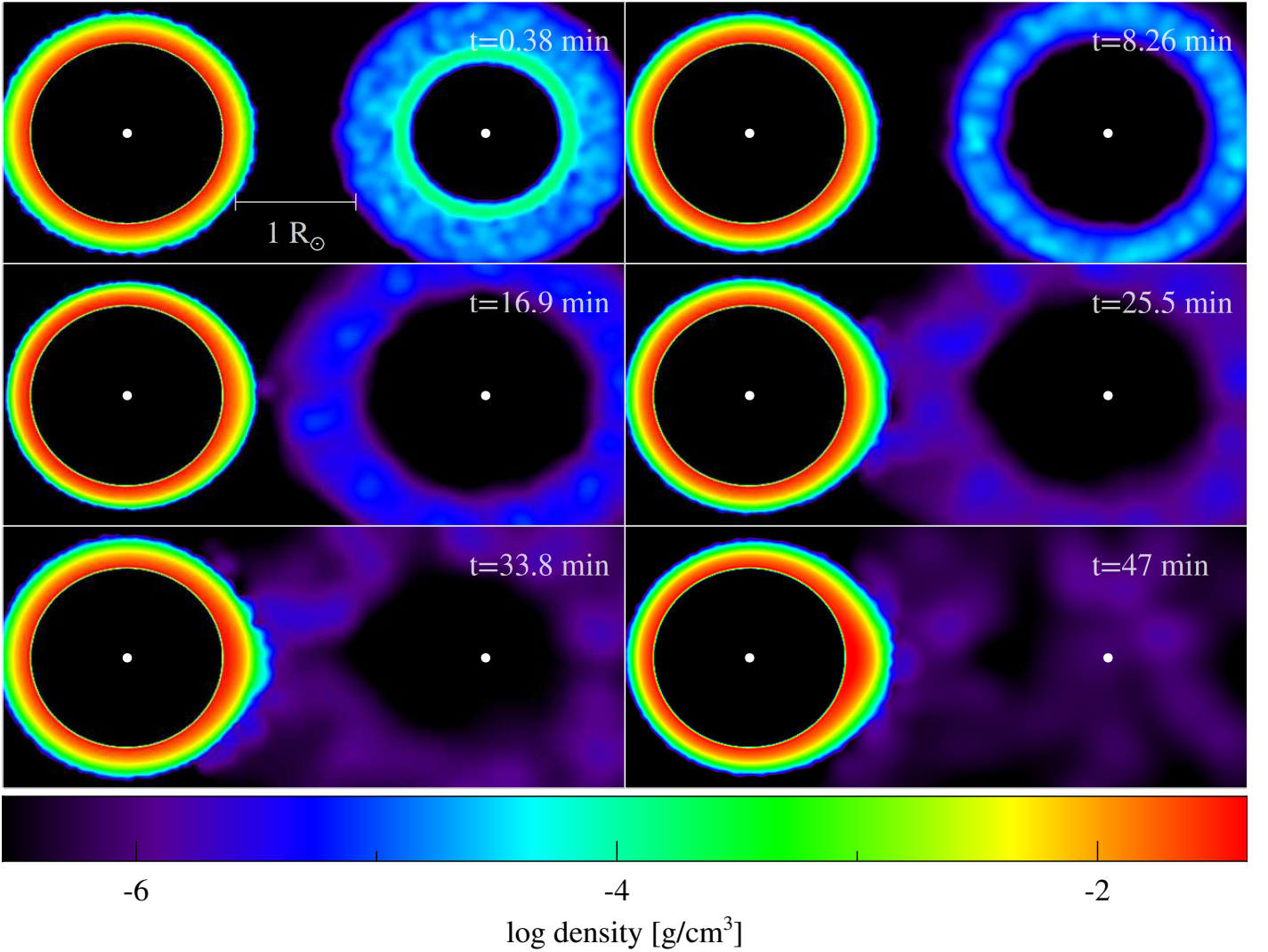}
\caption{Same as Fig. ~\ref{evolutionA}, but for density plots corresponding to Model G at different times. Note that in this model
the accretion disk does not get totally swept up in the impact with the nova ejecta. Snapshots and movies have been generated by means of the visualisation 
software SPLASH (Price 2007).}
     \label{evolutionG}
      \end{figure*}

Two values for the mass of the nova ejecta, $M_{\rm ejecta}$, have been considered to analyze the effect of this parameter. 
As shown in Table 2, a comparison between our fiducial Model A (characterized by $M_{\rm ejecta}$ = $5.1 \times 10^{-4}$ $M_\odot$) and Model E (with $M_{\rm ejecta}$ = $1.6 \times 10^{-3}$ $M_\odot$) reveals that, as expected, increasing the mass of the ejecta translates into larger masses gravitationally bound to the white dwarf ($\Delta M_{\rm WD}$) and to the main sequence companion ($\Delta M_{\rm MS}$), at the end of the simulations. In turn, the total amount of mass lost from the system, $\Delta M_{\rm esc}$, also increases. 
An identical pattern is found when comparing Models B and F (for which a maximum velocity of the nova ejecta of $V^{\rm max}_{\rm ejecta}$ = 800 km s$^{-1}$ has been assumed) and Models G and H ($V^{\rm max}_{\rm ejecta}$ = 1200 km s$^{-1}$, but twice the mass of the disk compared to Model A), with an exception: whereas the accretion disk gets fully disrupted in Models A, B, E, F, and H, it survives the collision with the ejecta in Model G (Fig. ~\ref{evolutionG}). This can be understood from the ratios $M_{\rm ejecta}$/$M_{\rm disk}$ adopted in the different models: the lowest value,  
$M_{\rm ejecta}$/$M_{\rm disk} \sim$ 13, corresponds to Model G, which suggests that only disks with masses much lower than the ejecta undergo total disruption. 
The fact that the disk gets disrupted in Model H but not in Model G affects the dynamics of the system and results in a moderately larger amount of mass that remains bound
to the white dwarf in the former. Except for such peculiar case, the fraction of the overall mass available (i.e., ejecta plus disk) that escapes the binary system (or remains bound to the main sequence or to the white dwarf) does not depend much on the choice of the mass of the nova ejecta (see Table 2). Note, indeed, that the fraction of mass leaving the system increases from 59\% to 89\%, while the fraction that remains bound to the white dwarf drops from 31\% to 8\%, when comparing Models G and H. 
As reported for Model A, small amounts of mass, up to $1.4 \times 10^{-6}$ $M_\odot$, involving only a few SPH particles, are expelled (if any) from the outer main sequence layers in the interaction with the nova ejecta, with values increasing for larger nova ejected masses.

\subsection{Effect of the velocity of the ejecta}
Three values for the maximum velocity of the ejecta, 
$V^{\rm max}_{\rm ejecta}$, representative of classical nova systems (Gehrz et al. 1998), 
have been adopted to analyze the influence of this parameter: 800 km s$^{-1}$, 1200 km s$^{-1}$, and 3000 km s$^{-1}$. 
Comparison between Models A and B reveals that an increase in the velocity of the ejecta yields larger ejected masses from the binary system, 
while reducing the amount of mass that remains gravitationally bound, either to the white dwarf or to the main sequence companion. 

The fraction of nova ejecta plus disk mass that escapes the binary system (or remains bound to the main sequence or to the white dwarf) follows exactly the same trend. 
 However, in sharp contrast to the results reported in Section 3.2, the specific fractions depend 
significantly on the values adopted for the velocity of the ejecta (Table 2). For instance, the fraction of mass leaving the system increases from 60\% to 89\%, while the fractions that remain bound to the white dwarf or to the main sequence drop from 30\% to 8\%, and from 10\% to 4\%, respectively, when comparing Models D and I. 
Similar patterns are observed regardless of whether the disk gets disrupted or not (see, e.g., 
Models C, D and I, Models E and F, and Models G and J, for which different combinations of masses of the ejecta and disk have been adopted). 
Larger velocities for the nova ejecta drive, as expected,  larger (but always tiny) amounts of mass lost by the main sequence companion, with a maximum value of $2.4 \times 10^{-6}$ $M_\odot$ achieved in Model I.

It is worth mentioning that while the accretion disk is disrupted in Models A, B, E and F, it survives the impact with the ejecta in all models characterized by moderately low 
$M_{\rm ejecta}$/$M_{\rm disk}$ ratios (e.g., Models C, D, I, and G, with $M_{\rm ejecta}$/$M_{\rm disk}$ ranging between 5.5 and 13). Note, however, that Model J, characterized by 
$M_{\rm ejecta}$/$M_{\rm disk}$ = 13, results in disk disruption too. This is caused by the large kinetic energy and momentum carried by the impinging ejecta, which in this particular model expands with a maximum velocity of 3000 km s$^{-1}$.

\subsection{Effect of the mass of the accretion disk}
The mass of the accretion disk that orbits around the white dwarf is a poorly constrained quantity. 
Accordingly, a series of disks with different masses, constructed in the framework of 
the Shakura \& Sunyaev model, have been considered. 
Comparison between Models A and C, for which two different values of the 
mass of the disk, $M_{\rm disk}$, have been adopted ($2 \times 10^{-5}$ and 
$10^{-4}$ $M_\odot$, respectively) reveals, as mentioned before, that the 
lower the mass of the disk, the larger the probability
of disruption. Hence, while the disk in Model A gets disrupted by the nova 
blast, it survives the impact in Model C. 
The same pattern is found for Models B and D, characterized by a lower 
expansion velocity of the nova ejecta (800 km s$^{-1}$),
and Models K and G, for which a different geometry of the disk, with 
a larger $H/R$ ratio, was assumed.
Increasing the mass of the accretion disk reduces in turn the amount of 
mass lost by the binary system (and conversely, increases the amount of mass 
that remains gravitationally bound to the white dwarf and to the main 
sequence companion)\footnote{Note, however, that 
Models K and G result in nearly identical ejected masses from the binary systems. 
This may partially result from the different geometry of the disks adopted
and from the smaller range of
values for the masses (a factor of 2 in Models G - K, versus a factor of 5
in Models A - C and B - D).}. 

The fraction of nova ejecta plus disk mass that escapes the binary system (or remains bound to the main sequence or to the white dwarf) follows the same pattern, with values showing a clear dependence on the accretion disk mass.

\subsection{Effect of H/R}
So far, we have analyzed the interaction of the nova ejecta with accretion 
disks characterized by a height to radius ratio of $H/R  = 0.03$. 
However, observations increasingly support a dispersion in the value of  
$H/R$, within the range 0.03 -- 0.1 (see Maccarone 2014; Shafter \& Misselt 
2006; Knigge et al. 2000). 
The influence of this parameter on the dynamical properties of 
the system has been analyzed by means of Model K, in which a ratio of $H/R  = 0.06$ has been adopted for the accretion disk.
Comparison between Model K and our fiducial Model A suggests that an 
increase in $H/R$ has a similar effect as a reduction in the 
velocity of the expanding nova ejecta: here, a larger $H/R$ results 
in an increase of the effective impact cross-section between disk and 
ejecta. This decelerates a larger fraction of the ejecta,
and as a result, the amount of mass gravitationally bound to the white dwarf 
and to the main sequence companion increases, while the overall mass lost 
from the binary system decreases. 
The fraction of nova ejecta plus disk mass that escapes the binary system (or remains bound to the main sequence or to the white dwarf) follows exactly the same trend. A thorough comparison between Models A and K reveals that the specific fractions depend significantly on the adopted $H/R$ ratio (Table 2). 
  
\section{Chemical pollution of the secondary star}

One of the possible outcomes of the dynamical interaction between the nova ejecta and the 
main sequence companion is the pollution of the secondary, enhancing the metal content of its outer layers.
The degree of contamination induced by the impact with the nova ejecta can be estimated from the overall number of particles gravitationally 
bound to the main sequence star (see, e.g., Lombardi et al. 2006). But this is by no means straight forward.
On the one hand, the models presented in this work follow the evolution of the binary system for about 3000 s, which corresponds 
to $\sim 0.1 P_{\rm orb}$. At this stage, a large number  of particles
 gravitationally bound to the secondary are still orbiting around in a corona that surrounds the star. 
Even though such particles will eventually fall into the star, it is difficult to anticipate how deep these particles will penetrate into its envelope.
Self-consistent calculations of these advanced stages would require a prohibitively intense computational effort
for many orbital periods to compute the corresponding infalling trajectories. 
And even if such numerical simulations would be feasible, a detailed account of the chemical profiles of the outer layers of the secondary would
require the use of more realistic initial models for the main sequence star (see, e.g., Sills \& Lombardi 1997) 
and the inclusion of important physical mechanisms that would be operating simultaneously (i.e., chemical diffusion, convection...). This is clearly
out of the scope of the present paper.
However, to illustrate the expected levels of chemical pollution, we have provide some crude estimates of the compositional changes in the outer layers of the secondary at different mass depths, for Model A. 
Assuming that all the gravitationally bound particles still in orbit will be incorporated and mixed with the outer $10^{-6}$ $M_\odot$ of the main sequence
companion, a mean metallicity of  $Z \sim 0.18$ is expected at those layers, in the hemisphere hit by the nova ejecta.
The expected level of contamination obviously decreases when considering deeper layers (i.e.,  $Z \sim 0.036$ for $10^{-5}$ $M_\odot$, and  $Z \sim 0.019$ for $10^{-4}$ $M_\odot$). 

A final issue involves the relevance of these results in the framework of systems 
with low-mass secondaries ($M \leq 2$ $M_\odot$), in which the presence
 of a convective envelope may wash out any trace of chemical pollution 
induced by the impact with the nova ejecta. 
This aspect has been addressed by Marks \& Sarna (1998) and 
Marks, Sarna \& Prialnik (1997), who analyzed the effect of re-accretion 
of material ejected during nova outbursts on the chemical evolution of 
the secondary, for binary systems with similar orbital periods and 
masses to those reported in this manuscript. 
The low-mass main sequence stars were evolved taking into account all 
major processes that may affect their surface composition: nuclear reactions, 
mass loss, convection, thermohaline mixing, and contamination with the 
nova ejecta. Using a control model, where no nova ejecta was incorporated 
into the secondary, they  reported elemental (e.g., C and N) and isotopic 
differences ($^{12}$C/$^{13}$C, $^{14}$N/$^{15}$N, $^{16}$O/$^{17}$O) on 
the surface layers of the secondary stars induced by the impact with the 
 ejecta. Therefore, one may expect some effect on the next nova cycle, once mass transfer onto the white dwarf component resumes, even in binaries 
with low-mass secondaries. This aspect, however, deserves in-depth analysis. 

\section{Conclusions}
Eleven 3D SPH simulations of the interaction between the nova ejecta, the accretion disk, and the stellar companion have been computed,  aimed at testing the influence of the 
different parameters (i.e., mass and velocity of the ejecta, mass and geometry of the accretion disk) on the dynamical and chemical properties of the binary system.
The main conclusions reached in this work are summarized as follows:
\begin{itemize}
\item We have investigated the conditions leading to the disruption of the accretion disk that orbits the white dwarf star. In 7 out of the 11 models computed, the disk gets fully disrupted and swept up.
In all these models, the disks are characterized by masses much smaller than that of the ejecta. Our simulations show that in
models with V-shaped disks with height-to-radius ratios of $H/R = 0.03$ and $M_{\rm ejecta}$/$M_{\rm disk} \leq 5.5$    
($M_{\rm ejecta}$/$M_{\rm disk} \leq 13$, for $H/R = 0.06$) the disk survives the impact with the nova blast.   
\item Small amounts of mass, up to $1.4 \times 10^{-6}$ $M_\odot$ are expelled from the outer main sequence layers in the interaction with the nova ejecta. No ejection is reported from 2 out of the 11 models computed.
\item An increase of the mass of the nova ejecta yields, in general, larger amounts of mass lost by the binary system, and larger masses gravitationally bound to the white dwarf and 
to the main sequence companion.  The fraction of the overall mass available (i.e., ejecta plus disk) that escapes the binary system (or remains bound to the main sequence or to the white dwarf) 
does not depend much on the choice of the mass of the nova ejecta. However, the dynamics of the system is influenced by disk disruption when the increase of the mass of the nova ejecta modifies the 
stability of the disk, moderately affecting the distribution of masses that remain
gravitationally bound to the white dwarf or to the main sequence, as well as the amount of mass lost from the system.
For instance, when comparing Models G (disk partially disrupted) and H (disk fully disrupted and swept up by the ejecta), the fraction of mass that leaves the binary system increases from 59\% to 89\%, while the fraction that remains bound to the white dwarf drops from 31\% to 8\%. 
\item An increase in the velocity of the ejecta results in larger ejected masses from the binary system, 
while reducing the amount of mass that remains gravitationally bound, either to the white dwarf or to the main sequence, 
 regardless of whether the disk gets disrupted or not. 
This results from the larger kinetic energy and momentum carried by the impinging ejecta when its velocity is increased.
The fraction of nova ejecta plus disk mass that escapes the binary system (or remains bound to the main sequence or to the white dwarf) follows exactly the same trend. The specific fractions depend much on the values adopted for the velocity of the ejecta.
The large kinetic energy and momentum carried by the ejecta in models with $V^{\rm max}_{ejecta} =$ 3000 km s$^{-1}$ can lead to disk disruption even for models characterized by relatively low $M_{\rm ejecta}$/$M_{\rm disk}$ ratios, as in Model J.
\item An increase of the mass of the accretion disk reduces the amount of 
mass lost by the binary system, and conversely, increases the amount of mass 
gravitationally bound to the white dwarf and to the main sequence companion.
For instance, when comparing Models A ($M_{\rm disk} = 2.04 \times 10^{-5}$ $M_\odot$) 
and C ($M_{\rm disk} = 9.28 \times 10^{-5}$ $M_\odot$), 
the fraction of mass that leaves the binary system decreases from 88\% to 67\%, 
while the fractions that remain bound to the white dwarf and to the main sequence star 
increase from 9\% to 25\%, and from 4\% to 8\%, respectively. 
This results from the smaller kinetic energy and momentum transferred to the disk particles per unit mass when the mass of the disk is increased, which
in turn reduces the probability of disk disruption by the nova blast.
\item An increase in the height-to-radius ratio of the disk has similar effects to a reduction of the 
velocity of the expanding ejecta: the larger effective impact cross-section between disk and 
ejecta slows down a larger fraction of the nova ejecta,  
which in turn increases  the mass gravitationally bound to the white dwarf 
and to the main sequence star, while reducing the overall mass lost 
by the binary system. 
\item A certain level of 
chemical contamination of the stellar secondary is induced by the impact with the nova ejecta (with a mean metallicity of  $Z \sim 0.18$ estimated at the outer $10^{-6}$ $M_\odot$ layers in the hemisphere hit by the ejecta, for Model A). This may have potential effects on the next nova cycle.
\end{itemize}

Since the problem is intrinsically three-dimensional  we cannot rely on 2D or axisymmetric approximations. A possible way to increase the resolution is to use a conical 3D computational domain, 
with the point-like white dwarf located at the vertex of the cone, so that only a fraction of the ejecta and disk, together with the full stellar secondary, are taken into account. The expected gain 
in resolution could reach a factor of $\sim$ 2, with $\sim$ 10 more particles in the conical computational domain, with respect to the simulations reported in this paper.

{\it Acknowledgements.} 
The authors would like to thank Ruben M. Cabez\'on, for many fruitful discussions and exchanges. 
This work has been partially
supported by the Spanish MINECO grant AYA2014--59084--P,
  and by the AGAUR/Generalitat de Catalunya grant 
SGR0038/2014.
SM is grateful to the South African National Research Foundation (NRF) for a research grant. 
This paper is dedicated to Enrique Garc\'\i a-Berro, who passed away in a tragic accident during the
revision of the manuscript.

\end{document}